\documentclass[a4paper,fleqn]{cas-dc}

\usepackage[numbers]{natbib}

\def\tsc#1{\csdef{#1}{\textsc{\lowercase{#1}}\xspace}}
\tsc{WGM}
\tsc{QE}

\begin{document}
\let\WriteBookmarks\relax
\def\floatpagepagefraction{1}
\def\textpagefraction{.001}


\shortauthors{T. Akaishi {\it et al.}}

\title [mode = title]{Precise lifetime measurement of $^4_\Lambda$H hypernucleus using in-flight $^4$He$(K^-, \pi^0)^4_\Lambda$H reaction}
\tnotemark[1]

\tnotetext[1]{Title footnote text}

%

\author[1]{T. Akaishi}
\author[2]{H. Asano}
\author[3]{X. Chen}
\author[4]{A. Clozza}
\author[4]{C. Curceanu}
\author[4]{R. Del Grande}
\author[4]{C. Guaraldo}
\author[3]{C. Han}
\author[5]{T. Hashimoto}
\cormark[1]
\ead{thashi@post.j-parc.jp}
\author[4]{M. Iliescu}
\author[1]{K. Inoue}
\author[6]{S. Ishimoto}
\author[2]{K. Itahashi}
\author[2]{M. Iwasaki}
\author[2]{Y. Ma}
\cormark[1]
\ead{y.ma@riken.jp}
\author[4]{M. Miliucci}
\author[2]{R. Murayama}
\author[1]{H. Noumi}
\author[7]{H. Ohnishi}
\author[9]{S. Okada}
\author[2]{H. Outa}
\author[4, 10]{K. Piscicchia}
\author[1]{A. Sakaguchi}
\author[2]{F. Sakuma}
\cormark[1]
\ead{sakuma@ribf.riken.jp}
\author[6]{M. Sato}
\author[4]{A. Scordo}
\author[1]{K. Shirotori}
\author[4,8]{D. Sirghi}
\author[4,8]{F. Sirghi}
\author[6]{S. Suzuki}
\author[5]{K. Tanida}
\author[1]{T. Toda}
\author[1]{M. Tokuda}
\author[2]{T. Yamaga}
\author[3]{X. Yuan}
\author[3]{P. Zhang}
\author[3]{Y. Zhang}
\author[11]{H. Zhang}
\affiliation[1]{organization={Osaka University},
  city={Osaka},
  postcode={567-0047}, 
  country={Japan}}
\affiliation[2]{organization={RIKEN},
  city={Wako},
  postcode={351-0198}, 
  country={Japan}}
\affiliation[3]{organization={Institute of Modern Physics, Chinese Academy of Sciences},
  city={Lanzhou},
  postcode={730000}, 
  country={China}}
\affiliation[4]{organization={Laboratori Nazionali di Frascati dell’ INFN},
  postcode={I-00044 Frascati},
  country={Italy}}
\affiliation[5]{organization={Japan Atomic Energy Agency},
  city={Tokai},
  postcode={319-1195}, 
  country={Japan}}
\affiliation[6]{organization={High Energy Accelerator Research Organization (KEK)},
  city={Tsukuba},
  postcode={305-0801}, 
  country={Japan}}
\affiliation[7]{organization={Tohoku University},
  city={Sendai},
  postcode={982-0826}, 
  country={Japan}}
\affiliation[8]{organization={Horia Hulubei National Institute of Physics and Nuclear Engineering (IFIN-HH)},
  city={Magurele},
  country={Romania}}
\affiliation[9]{organization={Chubu University},
  city={Kasugai},
  postcode={487-8501}, 
  country={Japan}}
\affiliation[10]{organization={CENTRO FERMI - Museo Storico della Fisica e Centro Studi e Ricerche Enrico Fermi},
  postcode={00184 Rome}, 
  country={Italy}}
\affiliation[11]{organization={Lanzhou University},
  city={Lanzhou},
  postcode={730000}, 
  country={China}}

\cortext[1]{Corresponding authors}

\nonumnote{}

\begin{abstract}
  We present a new measurement of the $^4_\Lambda$H hypernuclear lifetime using in-flight $K^-$ + $^4$He $\rightarrow$ $^4_\Lambda$H + $\pi^0$ reaction at the J-PARC hadron facility. We demonstrate, for the first time, the effective selection of the hypernuclear bound state using only the $\gamma$-ray energy decayed from $\pi^0$. This opens the possibility for a systematic study of isospin partner hypernuclei through comparison with data from ($K^-$, $\pi^-$) reaction. As the first application of this method, our result for the $^4_\Lambda$H lifetime, $\tau(^4_\Lambda \mathrm{H}) = 206 \pm 8 (\mathrm{stat.}) \pm 12 (\mathrm{syst.})\ \mathrm{ps}$, is one of the most precise measurements to date. We are also preparing to measure the lifetime of the hypertriton ($^3_\Lambda$H) using the same setup in the near future.
\end{abstract}


\begin{keywords}
 strangeness exchange reaction \sep $\pi^0$ tagging \sep hypernuclear weak decay lifetime
\end{keywords}

\maketitle
\section{Introduction}\label{intro}
Hypernuclear physics investigates the interaction between hyperons and nucleons in nuclei. Examples of significant findings in this field include the first direct evidence of nuclear mean field and the modification of nuclear structure by $\Lambda$ hyperon.\cite{Ch89,Li7L} The $\Lambda$ hypernuclear bound states have traditionally been produced through three reactions\cite{big_paper}: $^Z$A($K^-$, $\pi^-$)$^Z_\Lambda$A, $^Z$A($\pi^+$, $K^+$)$^Z_\Lambda$A and $^Z$A($\gamma^*$, $K^+$)$^{Z-1}_\Lambda$A. 
These reactions are known as the strangeness exchange reaction, associated strangeness production, and electromagnetic production, respectively.

Each of these three reactions has unique kinematic and dynamic characteristics. 
The strangeness exchange reaction is known for its small momentum transfer, which makes it a suitable choice for producing substitutional hypernuclear states by replacing one neutron with a $\Lambda$ hyperon. 
On the other hand, the isospin partner of strangeness exchange reaction,
$^Z$A($K^-$, $\pi^0$)$^{Z-1}_\Lambda$A, is experimentally challenging due to the detection of $\pi^0$ meson.
Despite the long-standing interest in producing isospin mirror hypernuclei and study isospin-dependent effects in hypernuclear structure with this reaction,
there has been no successful report except Ref. \cite{nms} and \cite{stopped_K}.
Ref. \cite{nms} measures the $\pi^0$ momentum from $(K_{stopped}^-, \pi^0)$ reaction and only limited missing mass resolution can be achieved;
Ref. \cite{stopped_K} detects $\pi^-$ from the mesonic weak decay of $^4_\Lambda$He($^4_\Lambda$H) from $^4$He$(K_{stopped}^-, \pi^{-(0)})$$^4_\Lambda$He($^4_\Lambda$H) reaction,
which suffers from large background induced by hadron reactions.
Both experiments show the difficulty for a systematic production of isospin mirror hypernuclei.
Alternatively, the $^Z$A($\gamma^*$, $K^+$)$^{Z-1}_\Lambda$A reaction can also convert a proton into a $\Lambda$ hyperon.
However, the strong spin-flip amplitude induced by the photon yields different quantum states from the strangeness exchange reactions.

J-PARC E73 collaboration propose a novel implementation for in-flight $^Z$A($K^-$, $\pi^0$)$^{Z-1}_\Lambda$A reaction by selecting fast $\pi^0$ mesons through the detection of high energy $\gamma$s in the forward direction. This novel method opens the possibility for a systematic study of isospin partner hypernuclei through comparison with data from the ($K^-$, $\pi^-$) reaction. Our setup is especially useful for coincidence measurements, such as hypernuclear mesonic weak decay and hypernuclear $\gamma$-ray spectroscopy. In this letter, we present the result of the first application of this method: the measurement of the $^4_\Lambda$H mesonic weak decay lifetime.

\section{J-PARC T77 experiment}

\begin{figure}[h]
  \begin{center}
    \includegraphics[width=8.0cm]{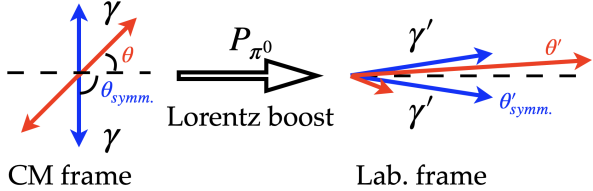}
    \caption{Illustration for the working principle of ($K^-$, $\pi^0$) event selection with $\gamma$ tagging in the forward direction.
      Blue and red lines show different boost effect for the decayed $\gamma$s.}
    \label{boost}
  \end{center}
\end{figure}

As the pilot run of our hypertriton ($_\Lambda^{3}$H) lifetime measurement experiment (J-PARC E73),
the J-PARC T77 experiment demonstrates the feasibility of in-flight $^{4}$He($K^-$, $\pi^0$)$_\Lambda^{4}$H reaction.
Instead of a full reconstruction of the $\pi^0$ momentum, we utilize the kinematics of the in-flight reaction to effectively select the events with hypernuclear bound states.
As illustrated in Fig. \ref{boost}, the emitted $\pi^0$ decays almost immediately into two $\gamma$s,
whose polar angle and energy in the laboratory frame is a function of the decay angle.
If we define $\theta_{symm.}$ as the symmetric opening angle of the two decayed $\gamma$s,
whose CM polar angle is perpendicular to the $\pi^0$ boost direction as illustrated by blue line on the left side of Fig. \ref{boost},
the decayed $\gamma$ with higher energy can always be detected by covering the very forward angle with $\theta' \leq \theta'_{symm.}$ (long red line)
because of stronger boost from the mother $\pi^0$ particle compared with $\theta' > \theta'_{symm.}$ case. 
This means that by tagging the high energy $\gamma$ in the forward angle, one can effectively select the fast $\pi^0$ in the forward direction from the ($K^-$, $\pi^0$) reaction.
Consequently, due to the small recoil momentum of the $\Lambda$ hyperon, these events are more likely to form the hypernuclear bound state.
The populated $^{4}_\Lambda$H hypernucleus can be identified through the coincidence measurement of the two-body mesonic weak decay,
$_\Lambda^{4}$H$\rightarrow$$^{4}$He+$\pi^-$.

In the J-PARC T77 experiment, we choose $K^-$ beam momentum at 1.0 GeV/$c$ to avoid background from $K^-$ in-flight decay and minimize the contribution from $\Sigma$ production. As a result, the $\pi^0$ emitted at 0 degrees from the $^{4}$He($K^-$, $\pi^0$)$_\Lambda^{4}$H reaction has a momentum of $\sim$0.9 GeV/c and the $\theta'_{symm.}$ is $\sim$7 degrees in the laboratory frame.
The recoil momentum of the populated $^4_\Lambda$H is obtained by using a differential cross section calculated by Harada. \cite{harada}
According to our stopping power simulation with SRIM package, the populated $^4_\Lambda$H stops within a few ps inside
the liquid $^{4}$He target, which has almost no effect for the $^4_\Lambda$H lifetime measurement.\cite{srim}
A technical challenge is to construct a suitable calorimeter to catch the high energy $\gamma$ ray in the very forward angle, which overlaps with the intensive meson beam. We solve this problem by constructing a Cherenkov-based electromagnetic calorimeter with PbF$_2$ crystals, which have a very short dead time and strong radiation robustness. We achieve a reasonably good energy resolution of 
$\sim5\%$ at 1 GeV energy deposit.\cite{panic21}

\begin{figure}[h]
  \begin{center}
    \includegraphics[width=8.0cm]{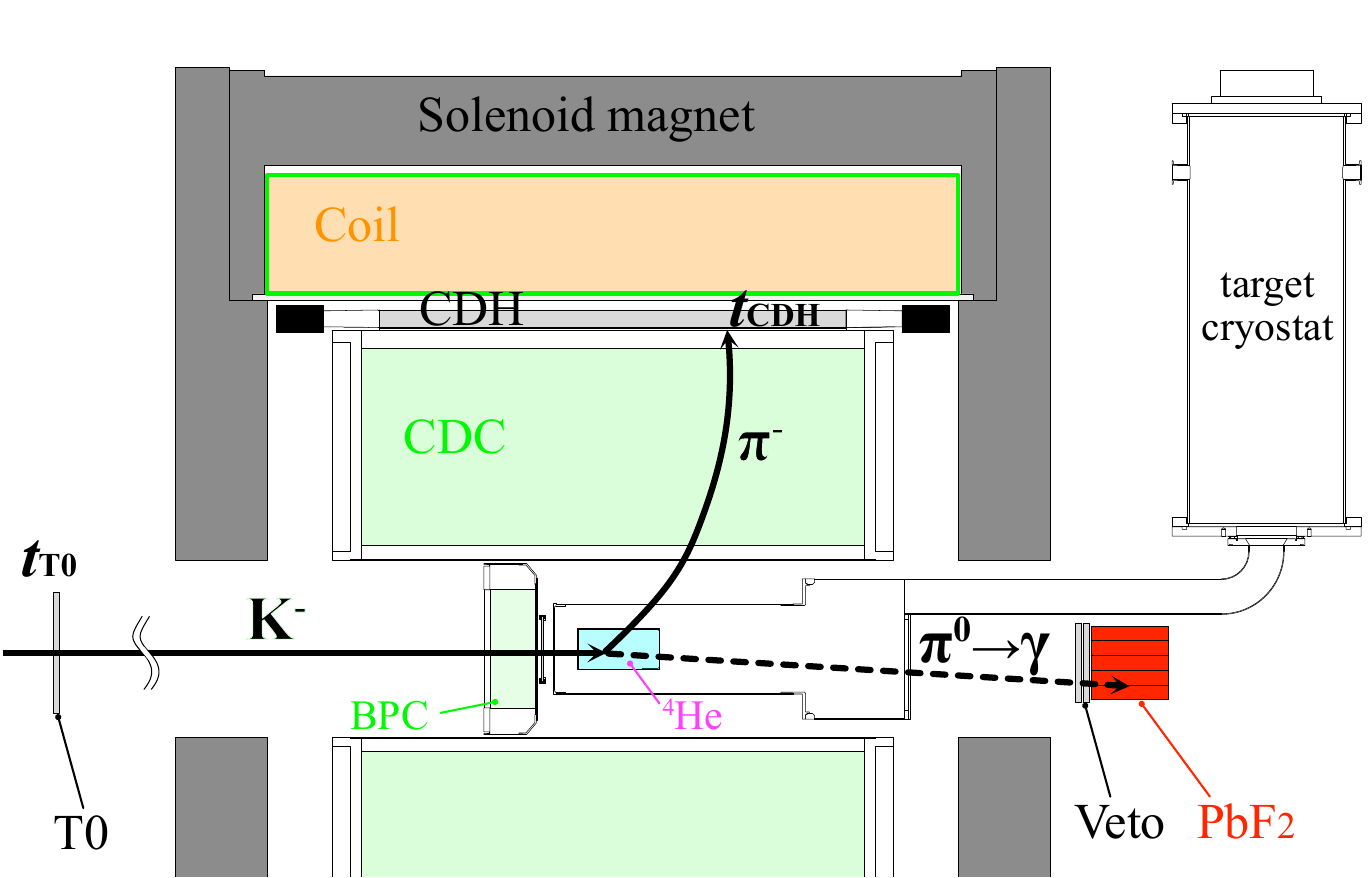}
    \caption{Schematic view of the J-PARC T77 experiment setup with liquid $^4$He target;
      high-energy $\gamma$ rays are tagged with PbF$_2$ calorimeter; Cylindrical Detector System(CDS) is the tracking device to capture decayed $\pi^-$ particle from $^{4}_\Lambda$H weak decay.}
    \label{setup}
  \end{center}
\end{figure}

The schematic view of the J-PARC T77 experiment setup is shown in Fig. \ref{setup}, which consists of a forward calorimeter, a Cylindrical Detector System (CDS), and a cryogenic system for liquid $^4$He target. The calorimeter is placed in the forward direction to detect the high energy $\gamma$ rays emitted from the $\pi^0$ decay. 
The CDS comprises a solenoid magnet, a Beam Profile Chamber (BPC), a Cylindrical Drift Chamber (CDC), and a Cylindrical Detector Hodoscope (CDH), which is used to detect the $\pi^-$ meson from the two-body mesonic weak decay $_\Lambda^{4}$H$\rightarrow$$^4$He+$\pi^-$ to identify the $_\Lambda^{4}$H hypernucleus.
The details of the CDS can be found in Ref. \cite{e15}.

The data taking for the J-PARC T77 experiment is conducted in June 2020 at the J-PARC hadron facility's K1.8BR beam line. Approximately $6\times 10^{9}$ of $K^-$ beams are irradiated onto a liquid $^4$He target, resulting in the successful identification of approximately 1.2$\times 10^3$ $_\Lambda^{4}$H hypernuclei using the $\pi^-$ decay spectrum. 

\section{Data analysis}

\subsection{Event selection}

To effectively select the $K^-$ beam, we use a threshold type Aerogel Cherenkov counter at the trigger level to veto $\pi^-$ and e$^-$ contamination in the beam. We also use the beam time-of-flight to further purify the $K^-$ beam in offline data analysis. To suppress background events with multiple charged tracks, such as multiple pion reactions, we 
require both single track in the CDC and single hit in the CDC inner layer.
The CDC has a vertex resolution of $\sim 1$ mm for r- and $\sim 7$ mm for z-direction. 
We also require consistency of $\phi$ angle between CDH hit segment and the CDC tracking.
Only events with $|\phi_{CDH}-\phi_{CDC}| < 7$ degrees are accepted to ensure the consistent between CDC tracking and CDH hit position.
The reaction vertex is reconstructed by combining a $K^-$ beam track measured by BPC and a $\pi^-$ track measured by the CDC.
A distance of closest approach (DCA) cut of $\le$ 5 mm is applied for vertex selection. 

\begin{figure}[ht]
  \begin{center}
    \includegraphics[width=\columnwidth]{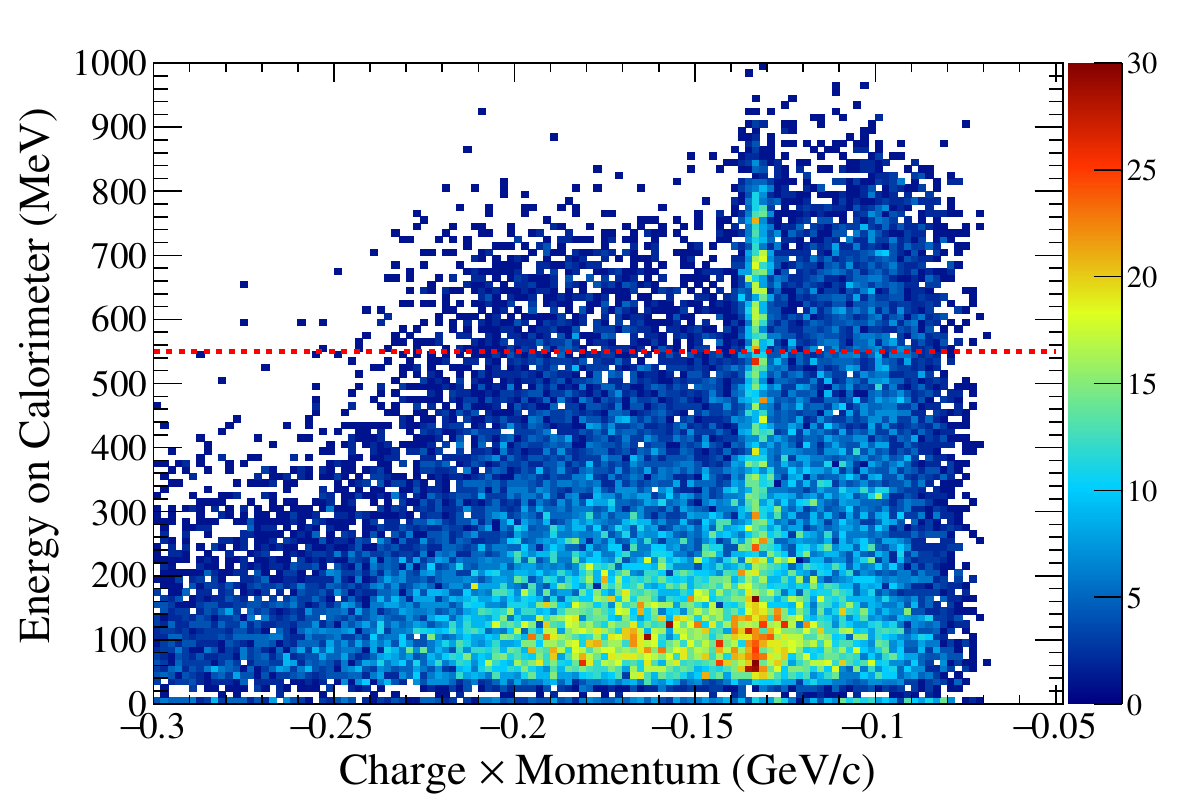}
         \caption{ Correlation between $\pi^-$ momentum and neutral particle energy deposit in the calorimeter. The red horizontal dotted line at 550 MeV shows the threshold for high-energy $\gamma$ tagging.}
    \label{calo_vs_pim}
  \end{center}
\end{figure}

\begin{figure}[ht]
  \begin{center}
    \includegraphics[width=\columnwidth]{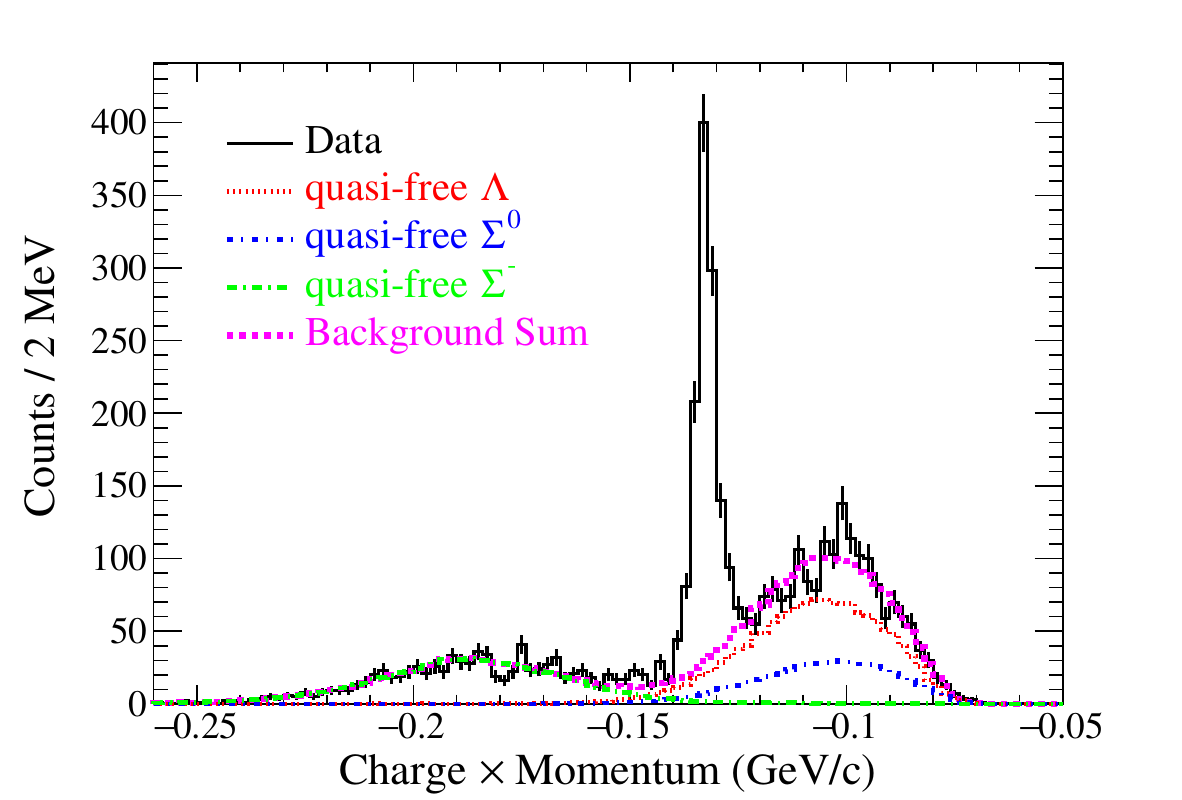}
    \caption{Momentum spectrum of the $\pi^-$. 
    The prominent $\pi^-$ peak around 132 MeV/c is from the two-body decay $_\Lambda^{4}$H$\rightarrow$$^{4}$He+$\pi^-$.
    The simulated background spectra from related quasi-free hyperon channels are overlaid.}
    \label{h4l_mom}
  \end{center}
\end{figure}

To effectively enhance the signal-to-noise ratio of the populated $^4_\Lambda$H hypernucleus,
the fast $\pi^0$ events in the forward direction are selected using large energy deposits in the calorimeter.
Fig. \ref{calo_vs_pim} shows the correlation between the $\pi^-$ momentum and energy deposit of the $\gamma$s
that have decayed from the $\pi^0$ after event selection. The effectiveness of our innovative $\pi^0$ tagging method is clearly demonstrated. 
Fig. \ref{h4l_mom} shows the final $\pi^-$ momentum spectrum after the high-energy $\gamma$ selection with $E_{\gamma}\ge 550$ MeV.
The $\pi^-$ produced from the two-body decay of $^4_\Lambda$H$\rightarrow$$^4$He+$\pi^-$ has a prominent peak with the correct momentum after energy loss correction.

The dominant background in the experiment is from the quasi-free $\Lambda$ and $\Sigma^{0,-}$ in-flight decay.
To reproduce the background events, we use the elementary reaction $K^- + N \to \Lambda /\Sigma^{0,-} + \pi^0$
convoluted with Fermi motion in Monte Carlo simulation and apply the same analysis as for the experimental data.\cite{fermi}
The experimental spectrum is fitted using the $\pi^-$ momentum distribution for each background processes and
calculated $^4_\Lambda$H production process to determine the relative amplitudes.\cite{harada}
As shown in Fig. \ref{h4l_mom}, excellent agreement between simulation and experimental data has been achieved,
which allows us to subtract the background time distribution and derive the $^4_\Lambda$H lifetime based on simulation.

\subsection{Timing analysis}

An important advantage of the J-PARC T77 (and also the future J-PARC E73) experiment is the capability to directly measure the $^4_\Lambda$H lifetime in time domain, which provides a unique opportunity to solve the hypertriton lifetime puzzle.\cite{hadron2021} In particular, the mesonic decay time of $^4_\Lambda$H can be obtained experimentally as 
\begin{eqnarray}
t_\mathrm{decay} = (t_\mathrm{CDH} - t_\mathrm{T0}) - t^\mathrm{calc.}_\mathrm{CDC} - t^\mathrm{calc.}_\mathrm{beam},
\end{eqnarray}
where $t_\mathrm{decay}$ stands for the decay time of the $^4_\Lambda$H event, $t_\mathrm{CDH} - t_\mathrm{T0}$ is the measured time difference between the CDH counter and T0 counter, $t^\mathrm{calc.}_\mathrm{CDC}$ is the calculated time of flight between the vertex and the CDH using the CDC track information, $t^\mathrm{calc.}_\mathrm{beam}$ is the calculated time of flight between T0 counter and the vertex using the momentum reconstructed with $K^-$ beam line spectrometer.

To optimize the time resolution, we perform time calibration and slewing correction for the CDH counter using prompt ($\pi^-$, $\pi^-$) events measured by the CDS, where the $\pi^-$ beam is mixed in and prescaled with the $K^-$ beam in trigger level. These events are selected in a similar manner to the ($K^-,\pi^0$) events, except that the requirement for the detection of a forward-neutral particle is removed. To cover our region of interest, only $\pi^-$ in the momentum range of ($-0.3$,$-0.1$) GeV/$c$ are used for the calibration.

\begin{figure}[t]
  \begin{center}
    \includegraphics[width=8cm]{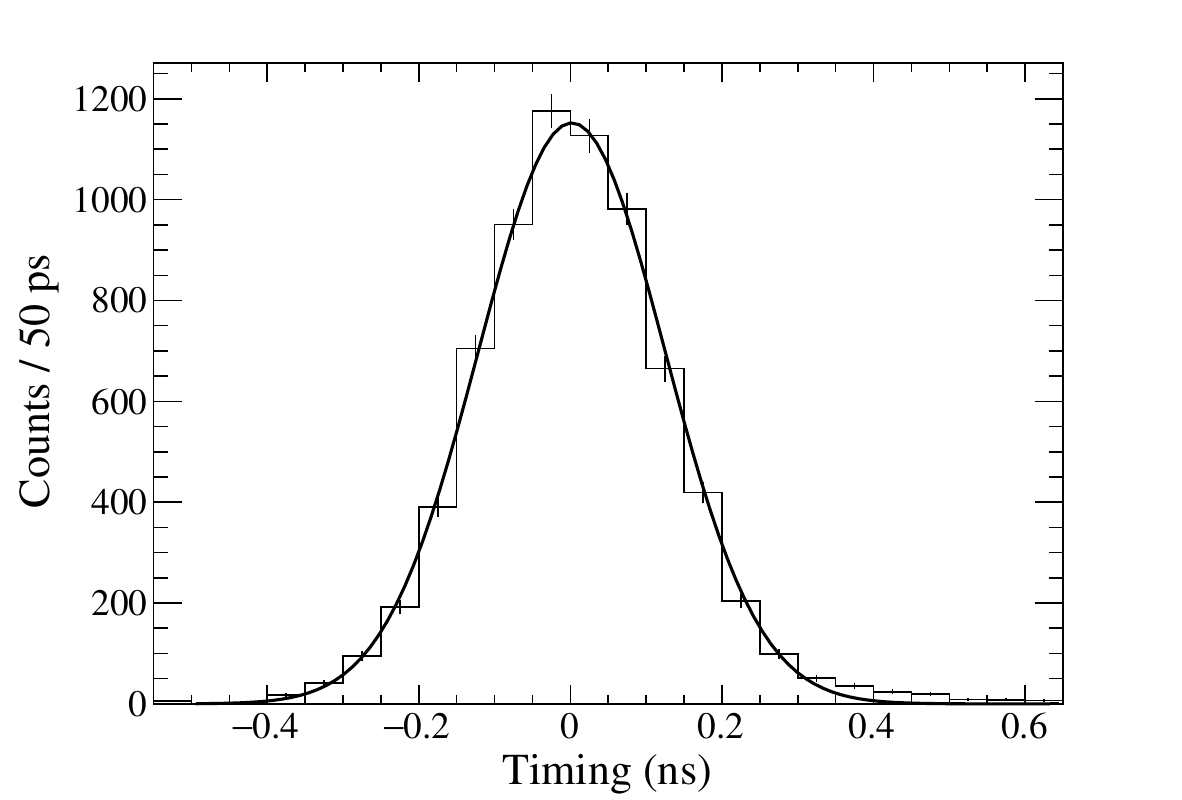}
    \caption{Time response function from the prompt ($\pi^-$, $\pi^-$) hadronic reaction.}
    \label{time_resp}
  \end{center}
\end{figure}

The time response function shown in Fig. \ref{time_resp} is obtained from the prompt ($\pi^-$, $\pi^-$) hadronic reaction at the $^4_\Lambda$H signal momentum range of ($-0.138$,$-0.124$) GeV/$c$. The response is well described by a Gaussian function with the standard deviation of $\sigma = 123 \pm 1$ ps.

\section{$^4_\Lambda$H lifetime result}
\begin{figure}[ht]
  \begin{center}
    \includegraphics[width=\columnwidth]{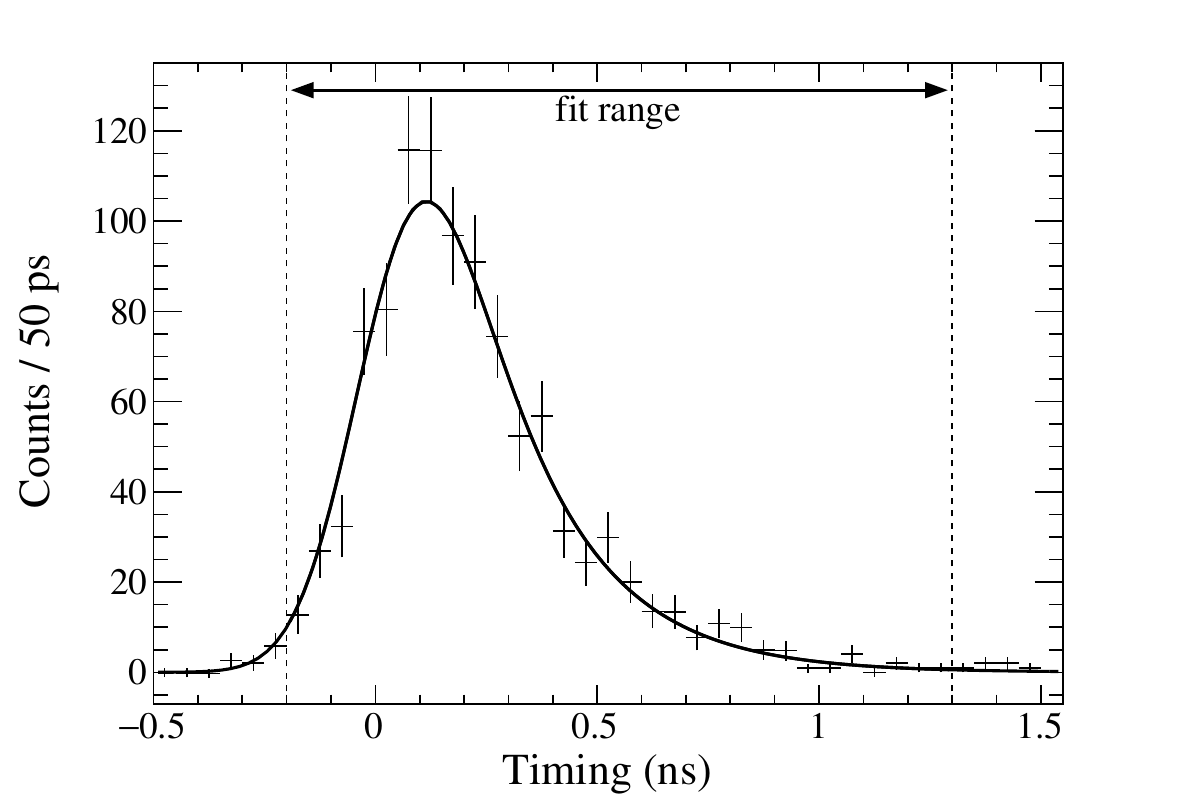}
    \caption{$^4_\Lambda$H lifetime fitting with time resolution function after background subtraction.}
    \label{h4l_fit}
  \end{center}
\end{figure}

The decay time spectrum at the $^4_\Lambda$H signal momentum range of ($-0.138$,$-0.124$) GeV/$c$ is obtained as shown in Fig. \ref{h4l_fit}.
Background contribution from the quasi-free hyperon processes are already subtracted based on the MC,
which is obtained by reproducing Fig. \ref{h4l_mom}.
The $^4_\Lambda$H lifetime is derived by fitting the time spectrum by an exponential distribution convoluted with time response function.
A binned minimum $\chi^2$ fit ($\chi^2/NDF=26.8/28$) gives
$\tau({^4_\Lambda}\textrm{H}) = 205.9 \pm 7.9\ \textrm{ps},$
where the quoted error is statistical only.

To ensure the accuracy of the measurement, it is important to carefully consider and quantify the sources of systematic uncertainty.
The main sources of systematic uncertainty for the measurement of the $^4_\Lambda$H lifetime are 
intrinsic bias in the experimental approach,
uncertainty in time calibration, 
uncertainty from $\gamma$ selection,
uncertainty in background subtraction, 
and uncertainty from the fitting process of the timing spectrum.
These uncertainties are summarized in Table \ref{tab:breakdown}. 

\begin{table}[h]
  \begin{center}
    \caption{Systematic uncertainties for $^4_\Lambda$H lifetime measurement.}
    \begin{tabular}{ m{0.65\linewidth} | m{0.15\linewidth} }
      \hline \hline
      Contribution & Value \\
      \hline
      Intrinsic bias of J-PARC T77 approach & $\pm$2 ps   \\
      Uncertainty from $\gamma$ selection &  $\pm$4 ps   \\
      Uncertainty of time calibration &  $\pm$7 ps   \\      
    Uncertainty of background subtraction &  $\pm$5 ps   \\
    Uncertainty in fitting process &  $\pm$7 ps   \\     
      \hline
      Total (quadratic sum) & $\pm$12 ps \\
      \hline\hline
    \end{tabular}
    \label{tab:breakdown}
  \end{center}
\end{table}





The intrinsic bias of the experimental approach refers to the systematic uncertainty in our method.
Specifically, the reaction vertex and decay vertex cannot be distinguished in our analysis.
The former can be displaced by a few millimeters with respect to the latter due to the forward recoil of the populated $^4_\Lambda$H hypernucleus. 
We evaluate such effect by the Monte Carlo simulation using a differential cross section calculated by Harada. \cite{harada}
Through a comparison of the input values and analysis results, we have determined that this intrinsic bias is controlled within $\pm$ 2 ps.

The selection of $\gamma$-ray energy can affect the background distribution from quasi-free hyperon decay.
We perform the full analysis procedure by varying the energy threshold by $\pm$50 MeV for both experimental data analysis and MC background evaluation.
The impact of this effect is found to be less than or equal to $\pm$4 ps.

The uncertainty in time calibration arises from run-by-run and segment-by-segment alignment, the dependence of hit position on our 79-cm long scintillation counter (CDH). We evaluate these uncertainties using the $(\pi^-,\pi^-)$ data. Additionally, a $0.5\%$ uncertainty in the $K^-$ beam momentum affects the calculation of $t^\mathrm{calc.}_\mathrm{beam}$. The overall uncertainty is confirmed to be within a tolerance of $\pm$7 ps.

The uncertainty in background subtraction arises from the unknown amount of background events and the timing distribution. To estimate this uncertainty, we intentionally scale the background events from quasi-free $\Lambda$ and $\Sigma$ in-flight decay by $\pm 2 \sigma$ from the best fit of our simulation. The ratio between quasi-free $\Lambda$ and $\Sigma^0$ is fixed in the nominal fitting based on the cross section in the literature since the momentum distribution of the two process are similar. We evaluate the effect of the uncertainty in this ratio by suppressing $\Lambda$ or $\Sigma^0$, respectively. We also employ a conventional side-band subtraction method, which is found to be consistent with the quoted systematic error.

The uncertainty in the fitting process is estimated by varying the time range and binning width. We also check the uncertainty by changing time resolution of the Gaussian response by $\pm 5$~ps, whose effect to the lifetime is found to be negligible.

By combining the fitting results with systematic uncertainties, we conclude the $^4_\Lambda$H lifetime as 
\begin{eqnarray}
\tau(^4_\Lambda \mathrm{H}) = 206 \pm 8 (\mathrm{stat.}) \pm 12 (\mathrm{syst.})\ \mathrm{ps},
\end{eqnarray}
which is consistent with previous results.\cite{outa, star2021}
The precision of $^4_\Lambda \mathrm{H}$ lifetime has been essentially improved by our experiment and other measurements\cite{star2021}, which allows a more reliable test for the $^4_\Lambda \mathrm{H}$ wave function especially for the repulsive core.\cite{outa_d}

\section{Summary and outlook}

The J-PARC T77 experiment successfully establishes a new method to produce the hypernuclei with in-flight $^Z$A($K^-$, $\pi^0$)$^{Z-1}_\Lambda$A reaction, which meets the long demanding request to study the isospin dependent hyperon-nucleon interaction. As the first application of this method, we derive the $^4_\Lambda$H lifetime with top class precision.\cite{outa, star2021}
Our next plan is to measure the hypertriton lifetime with the present method and solve the so called hypertriton lifetime puzzle.\cite{hadron2021}

\section*{Acknowledgements}
We would like to express our appreciation for the support from the J-PARC accelerator and hadron facility staffs.
We are also grateful for the inspiring discussion and kind help from Prof. T. Harada.
This project is partially supported by the MEXT Grants-in-Aid 17H04842, 19J20135, 21H00129, 18H05402, 26287057, and 22H04917.
We also acknowledge the support by International Partnership Program of the Chinese Academy of Sciences Grant No. 016GJHZ2022054FN.
T. H. was supported by a MEXT Leading Initiative for Excellent Young Researchers Grant. 




\printcredits

\bibliographystyle{elsarticle-num}

\bibliography{mybibfile}



\end{document}